\documentclass[preprint,showpacs,superscriptaddress,aps]{revtex4-1}
\usepackage{amssymb}
\usepackage{amsmath,amssymb,graphicx}
\usepackage{graphicx}
\usepackage{dcolumn}
\usepackage{bm}
\def\be{\begin{equation}}
\def\ee{\end{equation}}
\def\bea{\begin{eqnarray}}
\def\eea{\end{eqnarray}}

\def\lsim{\raise0.3ex\hbox{$\;<$\kern-0.75em\raise-1.1ex\hbox{$\sim\;$}}}
\def\gsim{\raise0.3ex\hbox{$\;>$\kern-0.75em\raise-1.1ex\hbox{$\sim\;$}}}

\usepackage{axodraw4j}
\usepackage{pstricks}
\begin{document}

\bigskip

\vspace{2cm}
\title{Lepton number violating four-body tau lepton decays}
\vskip 6ex
\author{G. L\'opez Castro}
\email{glopez@fis.cinvestav.mx}
\author{N. Quintero}
\email{nquintero@fis.cinvestav.mx}
\affiliation{Departamento de F\'\i sica, Centro de Investigaci\'on y de Estudios Avanzados, Apartado Postal 14-740, 07000 M\'exico D.F., M\'exico}

\bigskip

\bigskip

\begin{center}
\begin{abstract}
  We study the four-body $\tau^{\pm} \to \nu_{\tau}l^{\pm}l^{\pm}X^{\mp}$ decays where $l=e$ or $\mu$ and $X=\pi,\ K,\ \rho$ and $K^*$ mesons.  These decay processes violate the total lepton number ($|\Delta L|=2$ ) and can be induced by the exchange of Majorana neutrinos. We consider an scenario where these decays are dominated by the exchange of only one heavy neutrino which produces an enhancement of the decay amplitude via the resonant mechanism. Searches for these novel decay channels with branching fractions  sensitivities of $O(10^{-7})$ can provide constraints on the parameter space of the Majorana neutrinos which are stronger than the ones obtained from $\Delta L=2$ decays of charged pseudoscalar mesons.
\end{abstract}
\end{center}

\pacs{11.30.Fs, 13.35.Dx, 14.60.Pq, 14.60.St}
\maketitle
\bigskip

\section{ Introduction}

\bigskip

  Current evidence from oscillations experiments \cite{Fukuda:1998mi} allows to conclude that the involved neutrinos are very light massive particles and that their flavors are mixed. One of the most intriguing and still unsolved questions in particle physics is to elucidate if neutrinos are Dirac or Majorana fermions \cite{Mohapatra:2005wg}. The Majorana nature of neutrinos can be established in the simplest way through the observation of processes where the total lepton-number $L$ is violated by two units ($|\Delta L|=2$), a property that emerges from the non-invariance of the neutrino mass term \cite{Pontecorvo:1957qd,Gribov:1968kq,Bilenky:1980cx,Schechter:1980gr,Langacker:1986jv} under global phase transformations of Majorana fields. 

Up to now, most experimental efforts have focused on  searches of neutrinoless double beta nuclear decays \cite{Doi:1980yb}, which is by far the most sensitive $\Delta L=2$ channel. The non-observation of these decays \cite{KlapdorKleingrothaus:2000sn,Aalseth:2002rf,Arnaboldi:2008ds,Daraktchieva:2009mn,Rodejohann:2011mu} has provided very strong constraints on the existence of very light Majorana neutrinos, and has established direct upper bounds on the effective Majorana mass of the electron-neutrino $\langle m_{ee} \rangle$ at the sub-eV level \cite{Rodejohann:2011mu} (here we define $\langle m_{ll'} \rangle \equiv \sum_i U_{li}U_{l'i}m_i$, where $l,l'=e, \mu, \tau$ and $i=1,2,3$ labels the neutrino mass eigenstates; see Section II for notation). Direct bounds on other entries of the effective Majorana mass matrix are very poorly known \cite{Rodejohann:2011mu}, but indirect upper limits can be obtained by combining oscillation data \cite{GonzalezGarcia:2010er}, cosmological bounds \cite{Lesgourgues:2006nd,Sekiguchi:2009zs,Acero:2008rh,GonzalezGarcia:2010un} and tritium beta decay \cite{Otten:2008zz}. In turn, these indirect bounds on $\langle m_{ll'} \rangle$ can be used to predict other $\Delta L=2$ decays, such as same-sign dileptons produced in $\tau$ lepton or $K$, $D$ and $B$ meson decays. The predicted rates in this light Majorana neutrino scenario turn out to be extremely suppressed \cite{three-body,Ali:2006kw,Atre:2005eb} and beyond the sensitivities of current and future superflavor factories.

   A nice explanation for the very light scale of neutrinos can be found in the existence of additional heavy right-handed neutrinos via the see-saw mechanism \cite{Georgi:1974sy,Minkowski:1977sc,gellman, yanagida, glashow,Weinberg:1979sa,Mohapatra:1979ia}. As a remnant of lepton-number violating Majorana mass terms, the couplings of lepton charged currents can induce $\Delta L=2$ processes when expressed in the basis of  Majorana neutrino mass eigenstates. However, the exchange of very light or very heavy Majorana neutrinos in these decays  are strongly suppressed and usually also lead to unobservable rates \cite{three-body,Ali:2006kw,Atre:2005eb}. An alternative scenario is provided by the so-called {\it resonant mechanism} \cite{Atre:2009rg} which can produce large enhancements of the $\Delta L=2$ transition amplitudes if the masses of exchanged Majorana neutrinos are accessible to the energy scales of the physical processes. In this case, the non-observation of lepton number violating decays can be turned out into significant constraints on the mixings and masses of Majorana neutrinos.  Let us note that such heavy neutrinos in the range of a few keV to a few GeV can play an important role in cosmology and astrophysical processes \cite{Asaka:2005pn}, without conflicting neutrino oscillation data.  For instance, some extensions of the Standard Model incorporating right-handed singlet neutrinos provide a good candidate for dark matter in the form of a stable sterile neutrino in the range of a few keVs \cite{Asaka:2005pn}; also, such models contain additional heavier sterile neutrinos with masses of $O({1\ \rm GeV})$ which can explain the baryon asymmetry of the universe \cite{Asaka:2005pn}. Alternatively, Majorana neutrinos with masses in the range of a few hundred MeV to a few hundred GeV can be generated dynamically in extended technicolor model realizations of dynamical electroweak symmetry breaking \cite{intermediate}.

At low energies, the resonant enhancement scenario has been studied in several decays of pseudoscalar mesons and tau leptons.  The three-body decays $\tau^- \to l^+M_1^-M_2^-$, $K^+ \to \pi^-l^+l'^+$, $D^+ \to M^- l^+l'^+$, $B^+ \to M^- l^+l'^+$ have been considered in Refs. \cite{Atre:2009rg,Littenberg:1991ek,three-body,Zhang:2010um,Helo:2010cw,helo,Cvetic:2010rw} (charged conjugated modes are implied in all channels). In the case of the $\tau$ lepton decays some constraints can be derived on the product of two different mixing angles (for instance $|V_{\mu N}V_{\tau N}|$) as a function of the neutrino mass $m_N$ \cite{Atre:2009rg}, while the decays of pseudoscalar charged mesons allow to constrain also the individual mixing angles, for instance $|V_{lN}|$, as a function of the neutrino mass \cite{Atre:2009rg,helo}. Very recently, we have reported the first calculation of the four body decays $B^0 \to D^-l^+l^+\pi^-$ \cite{Quintero:2011yh}, which are expected to provide complementary constraints to the three-body decays of their charged counterparts. At higher energies, the production of same-sign dileptons at colliders \cite{Atre:2009rg,Dicus:1991fk,Han:2006ip,Kovalenko:2009td,Flanz:1999ah,Panella:2001wq,granada,almeida,yi} and in top quark decays \cite{Quintero:2011yh,top1,Si:2008jd} has also been considered in the literature. 

  Searches for $\Delta L=2$ three-body decays have been carried out by several experiments and the upper limits on the branching fractions can be found in Refs. \cite{pdg,He:2005iz,Rubin:2010cq,babar:2011hb}. New upper limits on branching fractions of lepton number violating decays of charged $B$ mesons have been reported recently: $(1)$ the Belle Collaboration \cite{belle2011} has obtained  $B(B^- \to l^-l^-D^+)\leq 10^{-6}$ ($l=e,\mu$) at the 95\% confidence level (CL); $(2)$ the BABAR Collaboration \cite{babar2012} has reported upper limits at the 90\% CL  for $B^+ \to h^- l^+l^+$ ($h=\pi/K$, $l=e/\mu$) of the order of a few times $10^{-8}$; $(3)$ the LHCb Collaboration has obtained results on the $B^- \to X^+\mu^-\mu^-$ decays ($X=D,\ D^*, \ D_s, \ \pi,\ K$) with upper bounds  ranging from $10^{-6}$ to $10^{-8}$ \cite{lhcb2011,lhcb2012}; in addition, an upper limit has been reported for the four-body decay $B(B^- \to D^0\pi^+\mu^-\mu^-)< 1.5\times 10^{-6}$ at the 95\% CL \cite{lhcb2012}.  Finally, searches for  $\Delta  L=2$ decays of the $\tau$ lepton have been reported by the Belle collaboration on six different $\tau^- \to l^+M_1^-M_2^-$ decay channels, with upper limits on branching ratios of the order of $10^{-8}$ \cite{:2009wc}. 

In the present paper we study the $\Delta L=2$  tau lepton decays $\tau^{\pm} \to \nu_{\tau}l^{\pm}l^{\pm}X^{\mp}$ (with $l=e$ or $\mu$, and $X=\pi,\ K,\ \rho$ or $K^*$ meson) within the scenario provided by the resonant Majorana mechanism. These decays allow to derive bounds on the $|V_{lN}|$ ($l=e\ \mu$) mixings, contrary to the case of three-body $\tau$ lepton decays which only allow to derive bounds on the product $|V_{lN}V_{\tau N}|$. Given the clean environment provided by $\tau$ lepton decays, these bounds on the Majorana neutrino mixings are free from hadronic uncertainties that are intrinsic to decays of pseudoscalar mesons. Therefore, these novel decay channels allow to derive constraints on the mixings that are complementary to those obtained from tau lepton and meson decays.  The large sample of $\tau$ lepton pairs ($\sim 10^{10}$) that are expected to be recorded at the superflavor factories \cite{superB,bevan2012}, makes very attractive the study of these lepton number violating processes.

\bigskip

\section{Charged currents of Majorana neutrinos and kinematics}

 The Feynman diagram corresponding to the $\tau^-(p)\to \nu_{\tau}(p_1) l^-(p_2)l^-(p_3)X^+(p_4)$ decays, where $X$ can be a pseudoscalar or a vector meson, is shown in Figure 1. The letters within brackets label the momenta of each particle. Following the definitions  given in Ref. \cite{Quintero:2011yh}, we can write the differential decay rate (in the rest frame of the decaying particle of mass $M$) in terms of five independent kinematical variables (see conventions in Figure 2):
\be
d\Gamma=\frac{X\beta_{12}\beta_{34}}{4(4\pi)^6M^3}\overline{|{\cal M}|^2}\cdot \frac{1}{n!}ds_{12}ds_{34}d\cos \theta_1 d\cos \theta_3 d\phi \ ,
\ee
where $s_{12}=(p_1+p_2)^2$ and $s_{34}=(p_3+p_4)^2$ denote the invariant masses of the 12 and 34 particles, while $(\theta_1,\ \theta_3,\ \phi)$ are angular variables defined in Figure 2 \cite{Quintero:2011yh,Pais:1968zz}. The $n$! factor in the denominator of Eq. (1) accounts for identical particles in the phase space, $\overline{|{\cal M}|^2}$ is the spin-averaged and  properly antisymmetrized (under exchange of identical leptons) squared amplitude, $\beta_{12}$ ($\beta_{34}$) is the velocity of particle 1 (particle 3) in the center of mass frame of particles 1 and 2 (3 and 4) and $X\equiv [(p^2-s_{12}-s_{34})^2-4s_{12}s_{34}]^{1/2}$. 

Similarly to previous studies \cite{top1,Atre:2009rg,Asaka:2005pn}, we add  $n$ right-handed singlets $N_{bR}$ $(b=1,2, \cdots n)$ fields to the usual three left-handed SU(2) lepton doublets $L_{aL}^T= (\nu_a,\ l_a)_L$, ($a=1,2,3$) of the Standard Model. In terms of the neutrino mass eigenstates obtained from the diagonalization of the Dirac and Majorana mass terms, we can write the charged current interactions of leptons as follows \cite{Atre:2009rg}:
\be
{\cal L}_{l}^{\rm ch}= - \frac{g}{\sqrt{2}} W^+_{\mu}
\left(\sum_{l=e}^{\tau}\sum_{m=1}^{3}U_{lm}\bar{\nu}_m\gamma^{\mu}P_Ll+  \sum_{l=e}^{\tau}\sum_{m=1}^{n}V_{lm}\overline{N^c_m}\gamma^{\mu}P_L l\right) + {\rm h. \ c.}
\ee
where $P_L=(1-\gamma_5)/2$ is the left-handed chirality operator,  $g$ is the $SU(2)_L$ gauge coupling, $\psi^c\equiv C\bar{\psi}^T$ is the charge conjugated spinor, and $U_{lm}$ ($V_{lm}$) denotes the mixings of light (heavy) neutrinos; the subscript $m$ refers to the basis of mass eigenstates obtained from the diagonalized Majorana mass term for neutrinos \cite{Atre:2009rg}:
\be
{\cal L}_m^{\nu}=-\frac{1}{2} \left(\sum_{m=1}^{3}m^{\nu}_m \overline{\nu_{mL}}\nu^c_{mR}+ \sum_{m=1}^{n}m^N_{m}\overline{N^c_{mL}} N_{mR} \right)+ {\rm h.\ c.}\ .
\ee

As in previous studies, we will assume that only one heavy neutrino with mass $m_N$ and charged current couplings $V_{lN}$ to leptons, dominates the decay amplitudes via the resonant enhancement mechanism. This scenario is useful to simplify the analysis of the parameter space, and it can be accomplished if the spectra of heavy neutrinos is such that only one of them falls in the mass region that is accessible in the decay under consideration ($m_l+m_{\pi} \leq m_N \leq m_{\tau}-m_l$ in the present case). 

\begin{figure}
  \includegraphics[width=8.5cm]{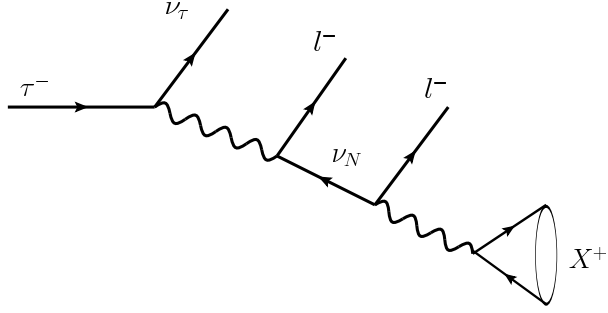}
  \caption{Feynman diagram  for the $\tau^-\to \nu_{\tau} l^-l^-X^+$ decay, where $X=\pi,\ \rho, K$ or $K^*$. The Majorana neutrino is denoted by $\nu_N$}\label{fig1}
\end{figure}

\begin{figure}
  \includegraphics[width=9cm]{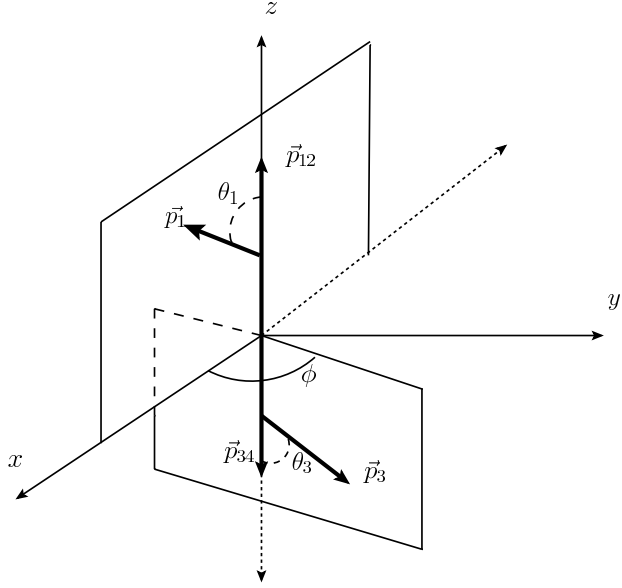}
  \caption{Kinematics of a generic four-body decay in the rest frame of the decaying particle, $\sum_{i=1}^{4}\vec{p_i}=0$. We have defined $\vec{p}_{ij}=\vec{p}_i+\vec{p}_j$, such that $\vec{p}_{12}+\vec{p}_{34}=0.$}\label{fig2}
\end{figure}

\bigskip

\section{Lepton number violation in four-body  $\tau$ lepton  decays}

  Following the convention of momenta defined in the previous section, we can write the (properly antisymmetrized) decay amplitude for the $\Delta L=2$ decays of the $\tau$ lepton as follows:
\be
{\cal M}= G_F^2V_{uq}V_{lN}^2m_N \bar{u}_{\nu_{\tau}}(p_1)\gamma^{\mu}P_Lu_{\tau}(p)\cdot \bar{u}(p_2)\left[{\cal P}_N(p_2)\gamma_{\mu}\gamma_{\nu}+ {\cal P}_N(p_3)\gamma_{\nu}\gamma_{\mu}\right]u^c(p_3)\left(V^{\nu}\right)\ .
\ee
Here, $G_F$ denotes de Fermi coupling constant, $V_{uq}$ (with $q=d$ or $s$) is the entry of the Cabibbo-Kobayashi-Maskawa matrix for the hadronic vertex, $V_{lN}$ is the neutrino mixing defined in Eq. (2) and $m_N$ denotes the mass of the heavy neutrino. In the case of two different charged lepton flavors in the final state, which we do not consider in this paper, we should  replace $V_{lN}^2 \to V_{eN}V_{\mu N}$. As it was stated before, we consider that only one heavy neutrino $N$ dominates the decay amplitude. The Lorentz-vector in Eq. (4) becomes $V^{\nu}=if_Pp_4^{\nu}$ when $X$ is a pseudoscalar meson and $V^{\nu}=f_Vm_V\epsilon^{\nu}(p_4)$ when $X$ is a vector meson. In our numerical evaluations, we will use the following values of the meson decay constants (all given in MeV units): $f_{\pi}= 130.4,  \ f_K=156.1$ from Ref. \cite{pdg}, and $\ f_{\rho}=216$ MeV , $f_{K^*}=205.4$ where obtained from the measured rates of $\tau \to V\nu_{\tau}$ decays quoted in \cite{pdg}. The lifetime of the $\tau$ lepton and the values of the quark mixing angles were taken also from \cite{pdg}.

In the expression for the decay amplitude we have introduced the factor
\be
{\cal P}_N(p_i)=\frac{1}{(Q-p_i)^2-m_{N}^2+im_{N}\Gamma_{N}}\ ,
\ee
where  $Q\equiv p-p_1=p_2+p_3+p_4$. In this expression, $\Gamma_N$ represents the decay width of the heavy neutrino; it allows to keep finite the amplitude when the heavy neutrino is produced on-shell, $(Q-p_i)^2=m_{N}^2$.   
For a given mass $m_N$ of the heavy neutrino, its decay width can be obtained by adding up the contributions of all its decay channels that can be opened at the mass $m_N$ \cite{Atre:2009rg}:
\be
\Gamma_N =\sum_f \Gamma(N\to f)\theta (m_N-\sum_{i} m_{f_i})\ ,
\ee
where $m_{f_i}$ in the argument of the step function are the masses of the final state particles in the neutrino decay channel $f$. The dominant decay modes of the neutrino in the range of masses that are relevant for the resonant $\tau$ lepton decays are induced by the exchange of $W^{\pm}$, see Eq. (2), and $Z^0$ gauge bosons: $N\to l^{\mp}P^{\pm}$, $\nu_l P^0$, $l^{\mp}V^{\pm}$, $\nu_l V^0$, $ l_1^{\mp} l_2^{\pm}\nu_{l_2}$, $\nu_{l_1}l_2^-l_2^+$, and $\nu_{l_1}\nu \bar{\nu}$, where $l,\ l_1,\ l_2=e,\ \mu$, and $P$ ($V$) denotes a pseudoscalar (vector) meson state. The expressions for the partial decay rates of these channels can be found in Appendix C of Ref. \cite{Atre:2009rg}.

 As it can be checked from Figure 4 in Ref. \cite{Quintero:2011yh}, the decay width $\Gamma_N$ varies between $10^{-20}$ GeV and (at most) $10^{-14}$ GeV for values of neutrino masses that are relevant for resonant $\tau^- \to \nu_{\tau} l^-l^-X^+$ decays. These numerical values are indeed upper limits and were obtained by assuming the bounds on the mixings of the heavy neutrino with the three charged leptons as reported in Ref. \cite{delAguila:2008pw}, namely:
\be
 |V_{eN}|^{2} \leq 3 \times 10^{-3}, \ \ |V_{\mu N}|^{2} \leq 3 \times 10^{-3}, \ \ |V_{\tau N}|^{2} \leq 6 \times 10^{-3}\ .
\ee
In other words, the neutrino decay width is so tiny that, for our purposes,  we can use the narrow width approximation,
\be
\lim_{\Gamma_N\to 0}{\cal P}_N(p_i)=-i\pi \delta \left((Q-p_i)^2-m_N^2\right)\ ,
\ee
to convert the five-dimensional integral in Eq. (1) into a four-dimentional one.  The branching ratios are then obtained by using the Montecarlo code VEGAS to perform numerically the four-dimensional integration. 

So far, no experimental searches have been reported for the $\tau^{\pm} \to \nu_{\tau}l^{\pm}l'^{\pm}X^{\mp}$ decays. With the large data sample of $\tau$ lepton pairs that are expected  at superflavor factories \cite{superB,bevan2012}, we may expect that sensitivities at or below the $10^{-7}$ level may be easily reached for the branching ratios of these decay channels. Just to illustrate the potencial of $\tau$ lepton decays to constrain the parameter space of the heavy neutrino, in Figure 3 (Figure 4) we show the calculated branching ratios for the di-electron (respectively, di-muon) channels as a function of the Majorana neutrino mass $m_N$ by using the upper bounds shown in Eq. (7).

\begin{figure}
  \includegraphics[width=13cm]{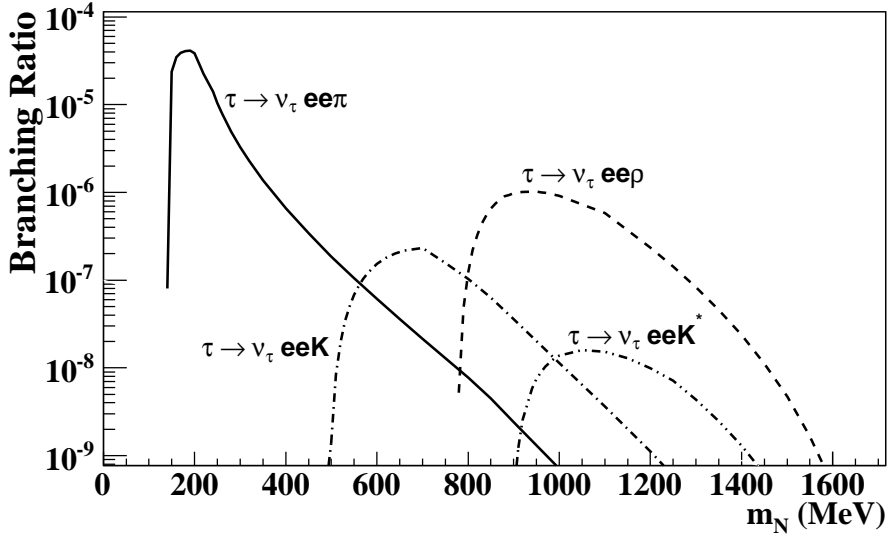}
  \caption{\small Branching fractions for $\tau^{\pm} \to \nu_{\tau}e^{\pm}e^{\pm}X^{\mp}$ decays as a function of $m_N$.}\label{fig3}
\end{figure}

\begin{figure}
  \includegraphics[width=13cm]{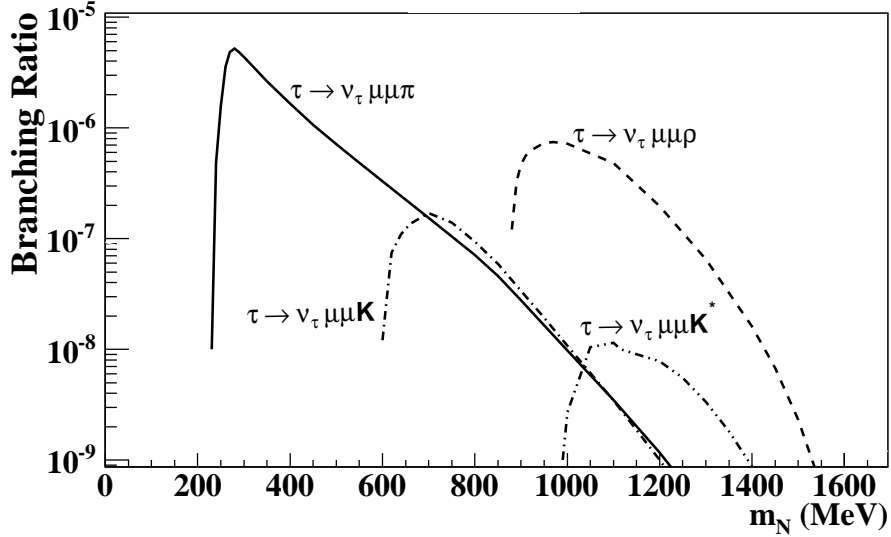}
  \caption{\small Same description as in Figure 3 for di-muon channels.}\label{fig4}
\end{figure}

  When upper bounds on the branching ratios of $\tau \to \nu_{\tau}Xll$ decays become available, we will will be able to get constraints in the $|V_{lN}|$ vs. $m_N$ plane, as done for example in Ref. \cite{Atre:2009rg}. These contraints are derived by noticing that, in the narrow width approximation, the dependence of the branching ratios upon the mixing angles is as follows:
\be
B(\tau^{\pm}\to \nu_{\tau}l^{\pm}l^{\pm}X^{\mp})\sim \frac{|V_{lN}|^2}{f_1|V_{eN}|^2+f_2|V_{\mu N}|^2} \cdot f_3\ ,
\ee
where $f_i$ ($i=1,\ 2,\ 3)$ depends upon the relevant coupling constants, phase-space integrals and the mass of the heavy Majorana neutrino.   In order to illustrate the constraints that can be gotten from the experimental searches, we will assume upper limits of  $O(10^{-7})$ for the branching ratios of different decay channels and we set $|V_{eN}|=|V_{\mu N}|$ \cite{note}. In Figures 5 and 6 we plot the exclusion regions (region above the plotted curves) in the $|V_{lN}|^2$ vs. $m_N$ plane for the di-electronic and di-muonic channels, respectively. 
\begin{figure}
  \includegraphics[width=13cm]{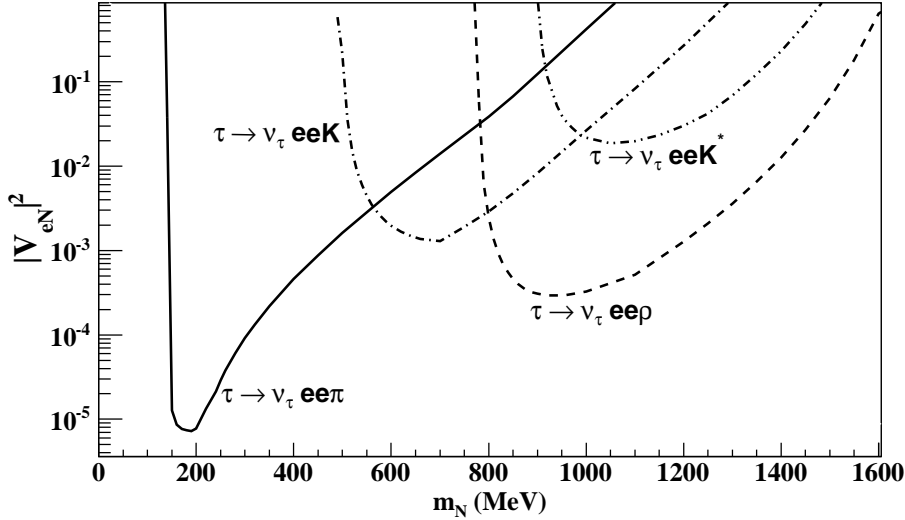}
  \caption{\small Exclusion regions in the $|V_{e N}|$--$m_N$ plane, by assuming upper bounds on $B(\tau^- \to \nu_{\tau}e^-e^-X^+)$ decays of order $10^{-7}$.}\label{fig5}
\end{figure}

\begin{figure}
  \includegraphics[width=13cm]{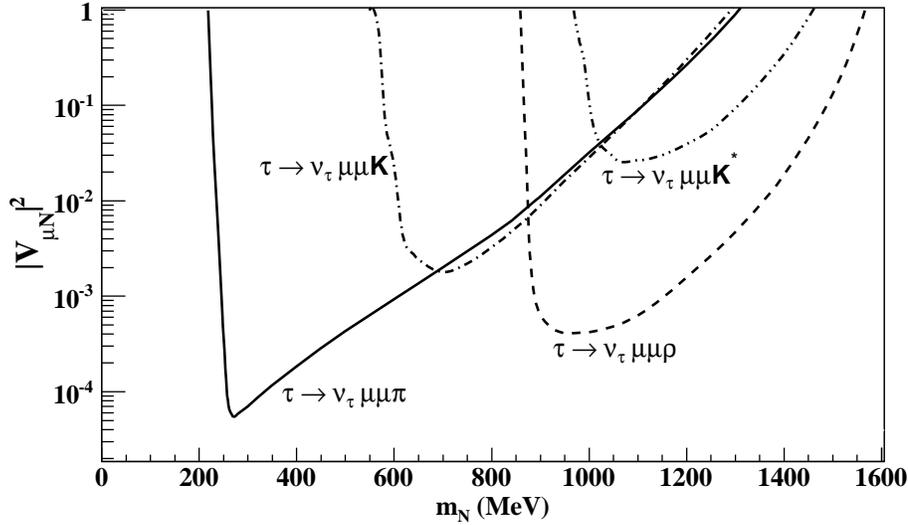}
  \caption{\small Exclusion regions in the $|V_{\mu N}|$--$m_N$ plane, by assuming upper bounds on $B(\tau^- \to \nu_{\tau}\mu^-\mu^-X^+)$ decays of order $10^{-7}$.}\label{fig6}
\end{figure}

As we can observe from Figures 5 and 6, both leptonic decay channels offer different sensitivities to the parameter space of the additional heavy neutrino. The searches of $\tau$ decays with different pseudoscalar and vector mesons in the final state would allow to constrain a larger and complementary region of the parameter space, although for the assumed $10^{-7}$ upper limits most of the excluded regions can be obtained from the strangeness-conserving channels. On the other hand,  a comparison of our results with Figures 9 and 11 of Ref. \cite{Atre:2009rg} shows that the $\tau$ lepton decays considered in this paper can provide stronger constraints on the $|V_{lN}|$ vs. $m_N$ parameter space than the ones coming from decays of $D$ and $B$ meson decays, at least in the region of neutrino masses where these decays overlap. The results of the present paper indicate that $\Delta L=2$ decays of the $\tau$ lepton can provide competitive constraints not only for the the product $|V_{lN}V_{\tau N}|$ of neutrino mixing angles, as is the case of three-body decays, but also over $|V_{lN}|$ without further theoretical uncertainties related to hadronic form factors or loop effects \cite{Ali:2006kw}.


\bigskip

\section{ Conclusions}

 In this paper we have studied the four-body $\tau^{\pm} \to \nu_{\tau}l^{\pm}l^{\pm}X^{\mp}$ decays, where $l=e$ or $\mu$ and $X=\pi,\ \rho,\ K$ and $K^{*}$ mesons. We consider a framework where the heavy Majorana neutrino that mediates these $\Delta L=2$ decays can enhance the decay amplitudes via the resonant mechanism and that the contribution of only one heavy neutrino dominates the decay amplitude. 

  We have found that these novel four-body decays, together with the three-body $\tau^{\pm} \to l^{\mp}M_1^{\pm}M_2^{\pm}$ decays previously studied by other authors, can provide a more complete set of constraints on the parameter space associated to the mass and mixings of the Majorana neutrino. One important advantage of these four-body  $\tau$ lepton decays is that they are free from the hadronic uncertainties associated to the decays of pseudoscalar charged mesons and depend only on well known decay constants of pseudocalar and vector mesons. By assuming experimental sensitivities of  $O(10^{-7})$ for branching ratios of different channels at superflavor factories, we find that the $\Delta L=2$ four-body  decays of $\tau$  leptons can provide constraints for mixing angles $|V_{lN}|^2 \sim 10^{-3}$ to $10^{-4}$, which are similar or better that the ones obtained from $B$ and $D$ meson decays.

\bigskip

\subsection*{Acknowledgements}

The authors are grateful to Conacyt (M\'exico)  for financial support.


\bigskip

\begin{thebibliography}{99} 
\bibitem{Fukuda:1998mi}
  Y.~Fukuda {\it et al.} [ Super-Kamiokande Collaboration ],
  Phys.\ Rev.\ Lett.\  {\bf 81}, 1562 (1998); 
  R.~Wendell {\it et al.} [ Kamiokande Collaboration ],
  Phys.\ Rev.\  {\bf D81}, 092004 (2010); 
  M.~Ambrosio {\it et al.} [ MACRO Collaboration ],
  Phys.\ Lett.\  {\bf B566}, 35 (2003).

\bibitem{Mohapatra:2005wg}
See for instance:  R.~N.~Mohapatra, S.~Antusch, K.~S.~Babu, G.~Barenboim, M.~-C.~Chen, A.~de Gouvea, P.~de Holanda, B.~Dutta {\it et al.},
  Rept.\ Prog.\ Phys.\  {\bf 70}, 1757 (2007).

\bibitem{Pontecorvo:1957qd}
  B.~Pontecorvo,
  Sov.\ Phys.\ JETP {\bf 7}, 172 (1958); 
  Sov.\ Phys.\ JETP {\bf 26}, 984 (1968).

\bibitem{Gribov:1968kq} 
  V.~N.~Gribov and B.~Pontecorvo,
  Phys.\ Lett.\ B {\bf 28}, 493 (1969).

\bibitem{Bilenky:1980cx}
  S.~M.~Bilenky, J.~Hosek and S.~T.~Petcov,
  Phys.\ Lett.\  B {\bf 94}, 495 (1980).

\bibitem{Schechter:1980gr}
  J.~Schechter, J.~W.~F.~Valle,
  Phys.\ Rev.\  {\bf D22}, 2227 (1980).

\bibitem{Langacker:1986jv}
  P.~Langacker, S.~T.~Petcov, G.~Steigman, S.~Toshev,
  Nucl.\ Phys.\  {\bf B282}, 589 (1987).

\bibitem{Doi:1980yb}
  M.~Doi, T.~Kotani, H.~Nishiura, K.~Okuda, E.~Takasugi,
  Phys.\ Lett.\  {\bf B102}, 323 (1981).

\bibitem{KlapdorKleingrothaus:2000sn}
  H.~V.~Klapdor-Kleingrothaus,  {\it et al.},
  Eur.\ Phys.\ J.\  {\bf A12}, 147 (2001).

\bibitem{Aalseth:2002rf}
  C.~E.~Aalseth {\it et al.} [ IGEX Collaboration ],
  Phys.\ Rev.\  {\bf D65}, 092007 (2002).

\bibitem{Arnaboldi:2008ds}
  C.~Arnaboldi {\it et al.} [ CUORICINO Collaboration ],
  Phys.\ Rev.\  {\bf C78}, 035502 (2008).
  
\bibitem{Daraktchieva:2009mn}
  Z.~Daraktchieva,
  Nucl.\ Phys.\  {\bf A827}, 495c (2009).


\bibitem{Rodejohann:2011mu} 
  For a recent review see: W.~Rodejohann,
  Int.\ J.\ Mod.\ Phys.\ E {\bf 20}, 1833 (2011)



\bibitem{GonzalezGarcia:2010er}
  M.~C.~Gonzalez-Garcia, M.~Maltoni, J.~Salvado,
  JHEP {\bf 1004}, 056 (2010). 

\bibitem{Lesgourgues:2006nd}
  J.~Lesgourgues, S.~Pastor,
  Phys.\ Rept.\  {\bf 429}, 307-379 (2006), and references cited therein.

\bibitem{Sekiguchi:2009zs}
  T.~Sekiguchi, K.~Ichikawa, T.~Takahashi, L.~Greenhill,
  JCAP {\bf 1003}, 015 (2010).
 
\bibitem{Acero:2008rh}
  M.~A.~Acero, J.~Lesgourgues,
  Phys.\ Rev.\  {\bf D79}, 045026 (2009).

\bibitem{GonzalezGarcia:2010un}
  M.~C.~Gonzalez-Garcia, M.~Maltoni, J.~Salvado,
  JHEP {}, 117 (2010).


 
 



\bibitem{Otten:2008zz}
  E.~W.~Otten and C.~Weinheimer,
  Rept.\ Prog.\ Phys.\  {\bf 71}, 086201 (2008).

 \bibitem{three-body} 
 A.~Ali, A.~V.~Borisov and N.~B.~Zamorin,
  Eur.\ Phys.\ J.\  C {\bf 21}, 123 (2001).

\bibitem{Ali:2006kw} 
  A.~Ali, A.~V.~Borisov and M.~V.~Sidorova,
  Phys.\ Atom.\ Nucl.\  {\bf 69}, 475 (2006)
  [Yad.\ Fiz.\  {\bf 69}, 497 (2006)].

\bibitem{Atre:2005eb}
  A.~Atre, V.~Barger and T.~Han,
  Phys.\ Rev.\  D {\bf 71}, 113014 (2005)  

\bibitem{Georgi:1974sy}
  H.~Georgi, S.~L.~Glashow,
  Phys.\ Rev.\ Lett.\  {\bf 32}, 438 (1974).


\bibitem{Minkowski:1977sc}
  P.~Minkowski,
  Phys.\ Lett.\  B {\bf 67}, 421 (1977).

\bibitem{gellman}M. Gell-Mann, P. Ramond and R. Slansky,
proceedings of the supergravity Stony Brook workshop, New York, 1979
(ed.s P. Van Nieuwenhuizen and D. Freedman, North-Holland, Amsterdam

\bibitem{yanagida}T. Yanagida, proceedings of the workshop on unified theories and baryon
number in the universe, Tsukuba, Japan 1979
 (ed.s. O. Sawada and A. Sugamoto, KEK Report No. 79-18, Tsukuba)
 
\bibitem{glashow}S.L. Glashow in``Quarks and Leptons'', Carg\'ese, 1979
 (ed.s M. L\'evy et al., North Holland 1980, Amsterdam)

\bibitem{Weinberg:1979sa}
  S.~Weinberg,
  Phys.\ Rev.\ Lett.\  {\bf 43}, 1566 (1979).


\bibitem{Mohapatra:1979ia}
  R.~N.~Mohapatra, G.~Senjanovic,
  Phys.\ Rev.\ Lett.\  {\bf 44}, 912 (1980).

\bibitem{Atre:2009rg}
  A.~Atre, T.~Han, S.~Pascoli and B.~Zhang,
  JHEP {\bf 0905}, 030 (2009) and references cited therein.
  
\bibitem{Asaka:2005pn} 
  T.~Asaka and M.~Shaposhnikov,
  Phys.\ Lett.\ B {\bf 620}, 17 (2005); 
  T.~Asaka, S.~Blanchet and M.~Shaposhnikov,
  Phys.\ Lett.\ B {\bf 631}, 151 (2005); M. Shaposhnikov and I Tkachev, Phys. Lett. B{\bf 639}, 414 (2006); M. Shaposhnikov, Nucl. Phys. B{\bf 763}, 49 (2007).

\bibitem{intermediate}
  T.~Appelquist, R.~Shrock,
  Phys.\ Rev.\ Lett.\  {\bf 90}, 201801 (2003);
  Phys.\ Lett.\ B {\bf 548}, 204 (2002); 
  T.~Appelquist, M.~Piai, R.~Shrock,
  Phys.\ Rev.\  {\bf D69}, 015002 (2004); 
  T.~Appelquist,
  Int.\ J.\ Mod.\ Phys.\  {\bf A18}, 3935 (2003).

\bibitem{Littenberg:1991ek}
  L.~S.~Littenberg and R.~E.~Shrock,
  Phys.\ Rev.\ Lett.\  {\bf 68}, 443 (1992).



\bibitem{Zhang:2010um}
  J.~M.~Zhang and G.~L.~Wang,
  Eur.\ Phys.\ J.\  C {\bf 71}, 1715 (2011).

\bibitem{Helo:2010cw}
 A. Ilakovac, B. A. Kniehl and A. Pilaftsis, Phys. Rev. {\bf D52}, 3993 (1995); A. Ilakovac and A. Pilaftsis, Nucl. Phys. {\bf B347}, 491 (1995); A. Ilakovac, Phys. Rev. {\bf D54}, 5653 (1996); V.~Gribanov, S.~Kovalenko, I.~Schmidt,
  Nucl.\ Phys.\  {\bf B607}, 355 (2001).

\bibitem{helo}
 J.~C.~Helo, S.~Kovalenko, I.~Schmidt,
    Nucl. Phys. B{\bf 853}, 80 (2011).

\bibitem{Cvetic:2010rw}
  G.~Cvetic, C.~Dib, S.~K.~Kang, C.~S.~Kim,
  Phys.\ Rev.\  {\bf D82}, 053010 (2010).

\bibitem{Quintero:2011yh} 
  D. Delepine, G.~L\'opez Castro and N. Quintero,
  Phys.\ Rev.\ D {\bf 84}, 096011 (2011).

\bibitem{Dicus:1991fk} 
  D.~A.~Dicus, D.~D.~Karatas and P.~Roy,
  Phys.\ Rev.\ D {\bf 44}, 2033 (1991).

\bibitem{Han:2006ip} 
  T.~Han and B.~Zhang,
  Phys.\ Rev.\ Lett.\  {\bf 97}, 171804 (2006)

\bibitem{Kovalenko:2009td}
  S.~Kovalenko, Z.~Lu and I.~Schmidt,
  Phys.\ Rev.\  D {\bf 80}, 073014 (2009).

\bibitem{Flanz:1999ah} 
  M.~Flanz, W.~Rodejohann and K.~Zuber,
  Phys.\ Lett.\ B {\bf 473}, 324 (2000)
  [Erratum-ibid.\ B {\bf 480}, 418 (2000)]; 
  Eur.\ Phys.\ J.\ C {\bf 16}, 453 (2000).

\bibitem{Panella:2001wq} 
  O.~Panella, M.~Cannoni, C.~Carimalo and Y.~N.~Srivastava,
  Phys.\ Rev.\ D {\bf 65}, 035005 (2002).

\bibitem{granada} 
F. del Aguila, J. A. Aguilar-Saavedra, and R. Pittau, J. Phys. Conf. Ser. {\bf 53}, 506 (2006); JHEP 0710, 047 (2007); F. del Aguila and J. A. Aguilar-Saavedra, Nucl. Phys. B{\bf 813}, 22 (2009).

\bibitem{almeida}
F. M. L. de Almeida, Jr. et al.,  Phys. Rev. D{\bf 75}, 075002 (2007).


\bibitem{yi}
Chien-Yi Chen, P. S. Bhupal Dev, e-print:arXiv:1112.6419 [hep-ph].



\bibitem{top1}
S.~Bar-Shalom, N.~G.~Deshpande, G.~Eilam, J.~Jiang and A.~Soni,
  Phys.\ Lett.\  B {\bf 643}, 342 (2006).

\bibitem{Si:2008jd}
  Z.~Si and K.~Wang,
  Phys.\ Rev.\  D {\bf 79}, 014034 (2009).

\bibitem{pdg} K. Nakamura,  et al. ( Particle Data Group), J. Phys. G.
\textbf{37 }, 075021 (2010).
 
\bibitem{He:2005iz}
  Q.~He {\it et al.} [ CLEO Collaboration ],
  Phys.\ Rev.\ Lett.\  {\bf 95}, 221802 (2005).
  
\bibitem{Rubin:2010cq}
  P.~Rubin {\it et al.}  [CLEO Collaboration],
  Phys.\ Rev.\  D {\bf 82}, 092007 (2010).

\bibitem{babar:2011hb} J.~M.~Link {\it et al.} [ FOCUS Collaboration ],
  Phys.\ Lett.\  {\bf B572}, 21 (2003);
J.~P.~Lees {\it et al.}  [The BABAR Collaboration],
  Phys.\ Rev.\ D {\bf 84}, 072006 (2011).

\bibitem{belle2011}
O. Seon {\it et al.} [Belle Collaboration], Phys. Rev. D{\bf 84}, 071106(R) (2011).

\bibitem{babar2012} 
J. P. Lees {\it et al.} [BABAR Collaboration], arXiv:1202.3650 [hep-ex].

\bibitem{lhcb2011}
R. Aaij {\it et al.} [LHCb Collaboration], arXiv: 1110.0730 [hep-ex]

\bibitem{lhcb2012}
R. Aaij {\it et al.} [LHCb Collaboration], arXiv: 1201.5600 [hep-ex]

\bibitem{:2009wc} 
  Y.~Miyazaki {\it et al.}  [BELLE Collaboration],
  Phys.\ Lett.\ B {\bf 682}, 355 (2010)


\bibitem{superB}
B. O'Leary {\it et al}, e-print arXiv:1008.1541 [hep-ex]; 
A. G. Akeyrod {\it et al}, http://belle2.kek.jp/physics.html.

 
\bibitem{bevan2012}
A. Bevan, J. Phys. G{\bf 39}, 023001 (2012).
 




\bibitem{Pais:1968zz}
  A.~Pais, S.~B.~Treiman,
  Phys.\ Rev.\  {\bf 168}, 1858-1865 (1968).
  

\bibitem{delAguila:2008pw}
  F.~del Aguila, J.~de Blas and M.~Perez-Victoria,
  Phys.\ Rev.\  D {\bf 78}, 013010 (2008).

\bibitem{note}
This assumption of $e-\mu$ universality is not compulsory. The exclusion region for the mixings and mass parameters  can be determined by scanning via a Monte Carlo sampling the allowed parameter space (see for example \cite{Atre:2009rg}). Here, we assume this approximation only to illustrate the potential of $\tau$ lepton decays to constrain these parameters.


\end{thebibliography}
\end{document}